\documentclass{iau} 
\usepackage{graphicx}

\title[Imaging the H-poor gas in PNe with large ADFs] %% give here short title %%
{Imaging the elusive H-poor gas in planetary nebulae with large abundance discrepancy factors}

\author[J. Garc\'{\i}a-Rojas et al.]   %% give here short author list %%
{Jorge Garc\'{\i}a-Rojas$^{1,2}$, Romano L.~M. Corradi$^{3,1}$, Henri M.~J. Boffin$^4$, Hektor Monteiro$^5$, David Jones$^{1,2}$, Roger Wesson$^6$, Antonio Cabrera-Lavers$^{3,1}$, \and Pablo Rodr\'{\i}guez-Gil$^{1,2}$}

\affiliation{$^1$Inst. de Astrof\'{\i}sica de Canarias, V\'{\i}a L\'actea, s/n, La Laguna, E-38205, Spain,  \\ email: {\tt jogarcia@iac.es} \\[\affilskip]
$^2$Dept. de Astrof\'{\i}sica, Universidad de La Laguna, La Laguna, E-38205, Spain \\ 
$^3$GRANTECAN, Cuesta de San Jos\'e s/n, E-38712 , Bre\~na Baja, La Palma, Spain\\ 
$^4$European Southern Observatory, Karl Schwarzschild Strasse 2, 85748 Garching, Germany \\
$^5$Univ. Fed. Itajub\'a, Av. BPS 1303-Pinheirinho, 37500-903, Itajub\'a, Brazil \\
$^6$Dept. of Physics \& Astronomy, University College London, Gower Street, London WC1E 6BT, UK \\
}

\pubyear{2016}
\volume{323}  %% insert here IAU Symposium No.
\jname{Planetary Nebulae: Multiwavelength probes os stellar and galactic evolution}
\editors{X.-W. Liu, L. Stanghellini \& A. Karakas, eds.}
\begin{document}

\maketitle

\begin{abstract}
The discrepancy between abundances computed using optical recombination lines (ORLs) and collisionally excited lines (CELs) is a major, unresolved problem with significant implications for the determination of chemical 
abundances throughout the Universe. In planetary nebulae (PNe), the most common 
explanation for the discrepancy is that two different gas phases coexist: a hot component 
with standard metallicity, and a much colder plasma enhanced in heavy elements. This dual nature is not predicted by mass loss theories, and direct observational support for it is still weak.
In this work, we present our recent findings that demonstrate that the largest abundance 
discrepancies are associated with close binary central stars. OSIRIS-GTC tunable filter 
imaging of the faint O~{\sc ii} ORLs and MUSE-VLT deep 2D spectrophotometry confirm that O~{\sc ii} ORL emission is more centrally concentrated than that of [O{\sc iii}] CELs and, therefore, that 
the abundance discrepancy may be closely linked to binary evolution.
\end{abstract}

\firstsection % if your document starts with a section,
\section{Introduction}

The abundance discrepancy problem was first discovered in the physics of gaseous nebulae more than 70 years ago (\cite[Wyse 1942]{wyse42}). It pertains to the fact that chemical abundances obtained from optical recombination lines (ORLs) are systematically larger than those obtained from collisionally excited lines (CELs) of the same ion. Generally, this discrepancy is parameterized by the abundance discrepancy factor (ADF), which for a given ion is defined as the ratio between ionic chemical abundances derived from ORLs and CELs. Photoionized gaseous nebulae exhibit different ADF values. While H\,{\sc ii} regions have moderate and nearly constant values ($\mathrm{ADF} \sim 2-3$, see \cite[Garc\'{\i}a--Rojas \& Esteban 2007]{garciarojasesteban07}), planetary nebulae (PNe) can reach values as large as 120 (\cite[Corradi et al. 2015]{corradietal15}).
%In photoionized gaseous nebulae, the behaviour of the ADF is different in H\,{\sc ii} regions, where moderate and fairly constabt values are found (ADF$\sim$2--3, see \cite[Garc\'{\i}a--Rojas \& Esteban 2007]{garciarojasesteban07}), and in planetary nebulae (PNe), where the ADF can reach values as high as 120 (\cite[Corradi et al. 2015]{corradietal15}). 
\cite{garciarojasesteban07} concluded that the physical origin of the abundance discrepancy in PNe with extreme ADF values should be different to that of H\,{\sc ii} regions and the bulk of PNe, where moderate values are found.

Regarding PNe, in the last few years our group has focused on the possible link between the large ADFs found in some PNe and the binary nature of their central stars (\cite[Corradi et al. 2015]{corradietal15}, \cite[Jones et al. 2016]{jonesetal16}). We have seen that it is relatively common to find extremely large ADFs in PNe with binary central stars that have experienced a common-envelope episode (post-CE PNe, e.\,g. \cite[Liu et al. 2006]{liuetal06}), reaching extreme values in the inner regions of several such objects (e.~g. in the central 7$''$ of Abell\,46, ADF$\sim$300). Our recent observations of the large ADF PN NGC\,6778 taken with the 10.4m Gran Telescopio Canarias (GTC) show that the spatial distribution of the O$^{2+}$ ions producing the O\,{\sc ii} ORLs does not match that of the O$^{2+}$ ions emitting the [O\,{\sc iii}] CELs; additionally, O\,{\sc ii} ORL emission is concentrated in the central parts of the nebula (\cite[Garc\'{\i}a--Rojas et al. 2016]{garciarojasetal16}). This is consistent with the presence of two distinct plasmas that are not well mixed, perhaps because they were produced in different ejection events. Our results are important because, for the first time, the evolution of close binary stars is clearly linked with the large ADFs in PNe. These results have encouraged us to acquire additional data in order to strengthen the link between the common-envelope process and the abundance discrepancy phenomenon. In particular, we have obtained high spatial resolution 2D spectroscopy with MUSE-VLT and direct imaging with a tunable filter using OSIRIS-GTC to study in detail the spatial distributions of both ORL and CEL emissions.

\section{Observations}

{\underline{\it GTC observations}}. We obtained narrow-band images of Abell\,46 using the blue tunable filter of the Optical System for Imaging and low Resolution Integrated Spectroscopy (OSIRIS) instrument at the 10.4m GTC (La Palma) on 20 September 2016. Due to tunable filter central wavelength is not constant across the size of Abell\,46 ($\sim$70$''$ diameter), a two-step scanning was required, centered at 4652  and 4656 \AA , respectively. We took 7 exposures of 1900 s in each configuration to reach the signal-to-noise ratio needed to detect the emission from the faint O\,{\sc ii} ORLs. For details on the observing and reduction technique, see \cite{garciarojasetal16}. 

{\underline{\it VLT observations}}. Observations were made with the Multi Unit Spectroscopic Explorer (MUSE) instrument mounted on VLT UT4 on 6 July 2016. The extended wide field mode was used, with a field of 60 $\times$ 60 $''$, 0.2 spaxel size and wavelength coverage 4650--9300 \AA\  at a mean spectral resolution of $\sim$2500, enough to isolate the emission of the brightest O\,{\sc ii} RLs. We observed 5 large ADF planetary nebulae. For each object, we made 5 long exposures (between 150 s and 1800 s, depending on target) to have enough S/N ratio to detect the faint O\,{\sc ii} RLs. We applied a dither pattern and a rotation of 90 degrees between the different exposures to avoid bad pixels and to get rid of any MUSE systematics. We also took two sky exposures to remove the sky-background emission from our final spectra.
We followed the standard reduction procedures described in \cite{walshetal16} using the instrument pipeline version 1.0 (\cite[Weilbacher et al. 2014]{weilbacheretal14)}).  Each datacube was completely reduced, sky subtracted and wavelength and flux calibrated. Only a single datacube for each object was used in the present work.

\section{Results}

\begin{figure}[!htbp]
\begin{center}
 \includegraphics[width=4.5in]{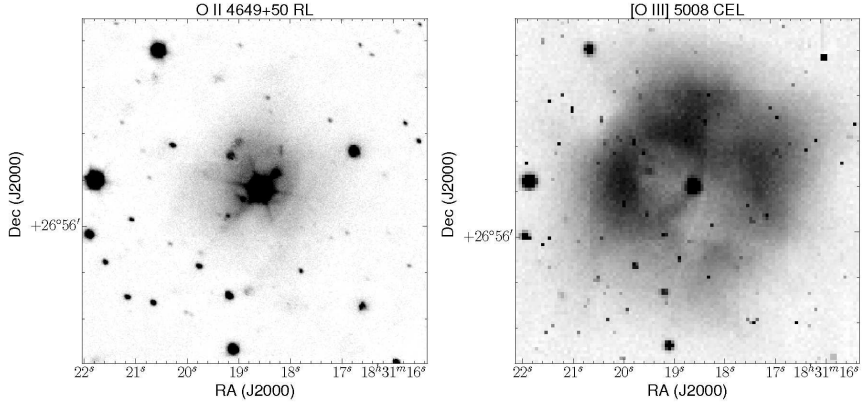} 
 \caption{OSIRIS-GTC tunable filter image of Abell\,46 in the O\,{\sc ii} 4649+50 ORLs (left), compared with an image taken with the 3.6m NTT in the [O\,{\sc iii}] 5007 CEL (right).}
   \label{fig1}
\end{center}
\end{figure}

Abell\,46 is the PN with the largest ADF ever measured in an integrated spectrum of a PN ($\sim$120, see \cite[Corradi et al. 2015]{corradietal15}); and has been shown to harbor a post-common-envelope central binary star (\cite[Afsar \& Ibanoglu 2008]{afsaribanoglu08}). In Fig.~\ref{fig1} we show the spatial distribution of the O\,{\sc ii} $\lambda$$\lambda$4649+50 ORLs emission (left) compared to the spatial distribution of the [O\,{\sc iii}] $\lambda$5007 CEL (right). Similar to what was observed in NGC\,6778 by \cite{garciarojasetal16} using the same technique, the emission from the ORLs is clearly produced closer to the binary central star than the CEL emission. This is consistent with the results of \cite{corradietal15} for Abell 46, who analyzed spatial emission profiles from long-slit spectroscopy (see their figure 4).

\begin{figure}[!htbp]
% \vspace*{-2.0 cm}
\begin{center}
 \includegraphics[width=4.8in]{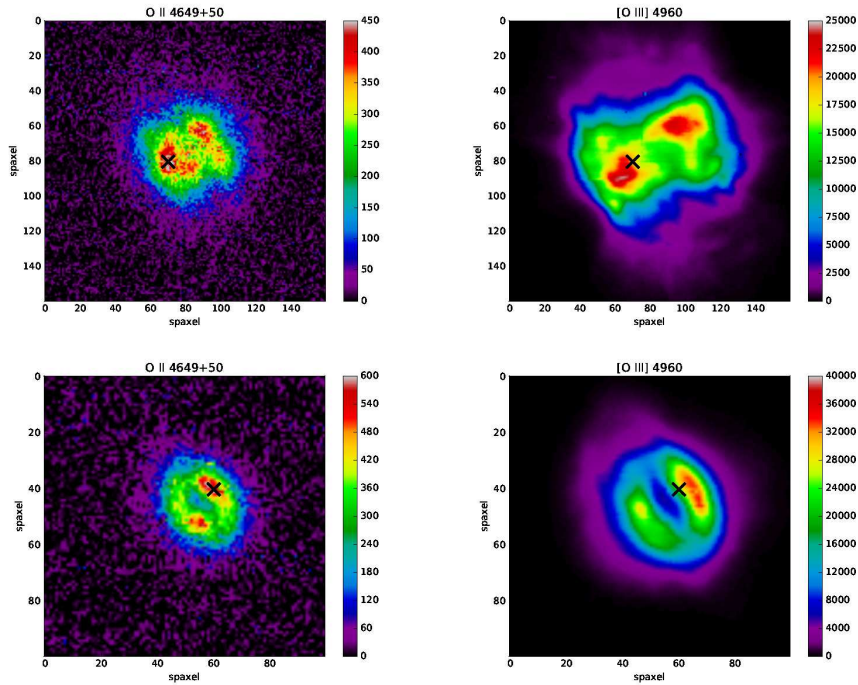} 
% \vspace*{-1.0 cm}
 \caption{Preliminary maps of the O\,{\sc ii} $\lambda\lambda$4649+50 ORL  (left panels) and [O\,{\sc iii}] $\lambda$4959 CEL emissions from two PNe of our sample: NGC\,6778 (ADF$\sim$20, top panels), and M\,1--42 (ADF$\sim$20, bottom panels). It is worth re-emphasising that the O\,{\sc ii} and [O\,{\sc iii}] emissions come from the same ion: O$^{2+}$. The ``x'' marks a reference spaxel.}
   \label{fig2}
\end{center}
\end{figure}

In Fig.~\ref{fig2} we show the continuum subtracted MUSE-VLT maps of the O\,{\sc ii} $\lambda$$\lambda$4649+50 ORL (left panels) and the [O\,{\sc iii}] $\lambda$4959 CEL (right panels) emissions for NGC\,6778 and M\,1--42. It is clear that in both objects the emission from the ORLs is more centrally concentrated than the emission coming from the strongest [O\,{\sc iii}] $\lambda$4959 CEL.
Both PNe have very similar ADFs of $\sim$20 (see \cite[Jones et al. 2016]{jonesetal16} and \cite[McNabb et al. 2016]{mcnabbetal16}). NGC\,6778 has a close binary central star that has undergone a common envelope phase (\cite[Miszalski et al. 2011]{miszalskietal11}). However, the central star of M\,1--42 has not been reported as binary, although its asymmetric morphology, large ADF and the behaviour found in this work suggests that it is an excellent candidate to search for binarity. 

We observed three additional objects: Hf\,2--2 a PN with one of the largest known ADFs ($\sim$80, \cite[McNabb et al. 2016]{mcnabbetal16}), which also hosts a close binary central star (\cite[Hillwig et al. 2016]{hillwigetal16}). NGC\,6153 and NGC\,7009 are very bright and well-studied PNe with relatively large ADFs ($\sim$10) but without known central binary stars (\cite[Liu et al. 2000]{liuetal00}, \cite[Fang \& liu 2013]{fangliu13}). For Hf\,2--2 there is also clear evidence of a more central concentrated emission of O~{\sc ii} ORLs relative to the emission of [O~{\sc iii}] CELs. However, given the brigthness of NGC\,6153 and NGC\,7009 and the relatively low S/N of O~{\sc ii} ORLs with respect to the continuum, a very careful continuum subtraction must be performed to clearly see the different spatial distributions of the O~{\sc ii} and [O~{\sc iii}] emissions.

Our observations have added a new, unexpected ingredient to understand the abundance discrepancy problem: large ADFs should be explained in the framework of close-binary evolution. A reasonable explanation related to binary evolution (nova ejecta, the presence of the common-envelope remnant or disk, etc.) should be invoked to explain the observed behavior.

\section{Acknowledgements}

Based on observations collected at ESO, Chile under proposal 097.D-0241, and at GTC, La Palma, Spain under proposal GTC3-16B. JGR thanks the IAU for a travel grant and Mexican grant UNAM-PAPIIT IN109614 for additional support of the research presented here. JGR also thank C. Esteban, M. Peimbert, S. Torres--Peimbert, C. Morisset and M. Pe\~na for support and discussions.

\begin{discussion}

\end{discussion}

\end{document}